\documentclass[pra,  longbibliography, reprint, %nofootinbib, 
amsmath,amssymb,amsfonts,floatfix]{revtex4-1}
\usepackage{braket}
\usepackage{graphicx}
\usepackage[hypertexnames=false,naturalnames=true]{hyperref}
\hypersetup{colorlinks=True}
\usepackage[all]{hypcap}
\usepackage{subfigure}
\usepackage{xcolor}
\usepackage{commath}

\usepackage{times}

\begin{document}
%\title{Photon Statistics Beyond Bunching and Antibunching in Waveguide QED}
%\title{Photon Statistics in a Waveguide: Beyond Bunching and Antibunching}
\title{Quantum Interference and Complex Photon Statistics in Waveguide QED}
\author{Xin H. H. Zhang}
\email{xin.z@duke.edu}
\author{Harold U. Baranger}
\email{baranger@phy.duke.edu}
\affiliation{Department of Physics, Duke University, P.O.\,Box 90305, Durham, NC 27708-0305, USA}

\date{February 2, 2018}
%\date{\today}

%%%%%%%%%----------------Abstract--------------------------------- %%%%%%%%%%%%%%%%%%%%%%%%
\begin{abstract}
%Photon bunching and antibunching are usually defined through the second-order correlation function $g^{(2)}(\tau)$. Here, 
We obtain photon statistics by using a quantum jump approach tailored to a system in which one or two qubits are coupled to a one-dimensional waveguide. Photons confined in the waveguide have strong interference effects, which are shown to play a vital role in quantum jumps and photon statistics. For a single qubit, for instance, bunching of transmitted photons is heralded by a jump that \emph{increases} the qubit population. We show that the distribution and correlations of waiting times offer a clearer and more precise characterization of photon bunching and antibunching. Further, the waiting times can be used to characterize complex correlations of photons which are hidden in $g^{(2)}(\tau)$, such as a mixture of bunching and antibunching.
\end{abstract}

\maketitle

%%%%%%%%%----------------Introduction--------------------------------- %%%%%%%%%%%%%%%%%%%%%%%%
\section{Introduction}
Photon statistics is one of the key ways to characterize non-classical light \cite{SrinivasOA1981, PaulRMP1982, LoudonBook2000, CarmichaelBook2003, CarmichaelBook2008}. One intuitively appealing aspect is whether the photons are \emph{bunched} (tending to arrive in bundles) or \emph{antibunched} (tending to arrive one-by-one). 
% One important aspect of photon statistics is the photon bunching/antibunching effect, which describes the correlations of photon arrivals. [HUB: the statistics of photon arrivals is the photon statistics, so I don't think this last phrase adds much.] For bunched photons, they tend to arrive as bundles; while for anti-bunched photons, they tend to arrive one-by-one. 
In the development of quantum optics, photon antibunching, which is forbidden for classical light, was used as a proof of the quantum nature of photons \cite{PaulRMP1982}. Now with the rapid development of quantum technologies, photon bunching/antibunching finds diverse applications such as the creation of strongly correlated photons \cite{ChangNPhoton2014, RoyRMP2016, GuPhysRep17}, design of new light sources \cite{ObrienNPhoton2009, MunozNPhoton2014, AharonovichNatPhoton2016}, and study of quantum many-body physics \cite{CarusottoRMP13, NohRPP2017, LeePRL2012, BardynPRL2012, ManzoniNatComm2017, FinkNatPhys2017}. To provide the requisite control, many of these applications involve photons in one-dimensional (1D) waveguides.

Photon bunching/antibunching is customarily defined in terms of the second-order correlation function $g^{(2)}(\tau)$ (the intensity-intensity correlation function normalized to the mean intensity). The simplest definition is that bunching (antibunching) occurs when $g^{(2)}(0)$ is larger (smaller) than $1$. That this definition is not sufficiently precise \cite{ZouPRA1990, CarmichaelBook2003} motivated other definitions, such as that bunching (antibunching) occurs when $g^{(2)}(\tau)$ is larger (smaller) than $g^{(2)}(0)$ \cite{LoudonBook2000}. However, this improved definition is ambiguous if $g^{(2)}(\tau)$ is structured or oscillating, as is very often the case in waveguide quantum electrodynamics (QED) \cite{ZhengPRA10, KocabasPRA2012, FangPRA15, RoyRMP2016, GuPhysRep17} because of the strong interference effects in one dimension. To characterize bunching/antibunching, then, more sophisticated photon statistics such as higher-order correlation functions are clearly needed \cite{CarmichaelPRA1989, PlenioRMP1998, MunozNPhoton2014}. Since quantum jumps can describe the detection of single photons emitted from a quantum system under continuous monitoring \cite{PlenioRMP1998, GardinerBook2004, CarmichaelBook2008, WisemanBook2014}, they offer a natural way to study the arrival times of photons and make the full photon statistics available. 

Confinement to one dimension has a profound effect on the quantum optical properties of a system because interference effects are much stronger and photonic fields do not decay with distance. Previous studies of waveguide systems \cite{RoyRMP2016, GuPhysRep17} have shown nonclassical photon statistics both theoretically and experimentally. A variety of techniques were used theoretically (not including quantum jumps) to find the correlation function $g^{(2)}(\tau)$ and the distribution of photon number within a pulse (Poisson distribution for a coherent state) (see for example Refs.\,\cite{ShenPRL07, *ShenPRA07, ZhengPRA2010, RoyPRL11, ZhengPRL13, PeropadreNJP13, LindkvistNJP14, RoyPRA14, LaaksoPRL14, FangPRA15, PletyukhovPRA2015}), and the full counting statistics of photons was found in the case of a single qubit \cite{PletyukhovPRA2015} though the connection to bunching/antibunching was not emphasized.  
Experimentally, non-classical deviations in the correlation function have been seen \cite{HoiPRL12,HoiNJP13,LangPRL11}. 
With regard to quantum jumps, jump operators that include photon interference have been used to describe the superposition of an output field and a coherent field in several situations \cite{CarmichaelBook2008, WisemanBook2014,CarmichaelPRL1993, PlenioRMP1998, GardinerBook2004, MirzaJOSAB2013, BaragiolaPRA2017,ManzoniNatComm2017}, including very recently photon detection at the output ends of a waveguide system \cite{BaragiolaPRA2017,ManzoniNatComm2017}. Quantum jumps have not been used previously, however, to study photon statistics in waveguides. 
Experimentally, the study of quantum jumps of a single quantum dot spin has been accomplished with a superconducting single photon detector and photon waiting times were measured \cite{DelteilPRL2014}. Recent progress in circuit QED experiments \cite{GirvinBook2014, QuantMicro2016, HadfieldBook2016}, including the observation of quantum trajectories \cite{WeberCRP2016} and single microwave photon detection \cite{ChenPRL2011, InomataNCommms2016, Sathyamoorthy2CRP016}, also render the experimental observation of microwave photon arrivals possible in the near future.
While many issues have clearly been investigated in waveguide QED, the role of strong interference in photon statistics and the existence and characterization of more complex photon statistics 
%are not very clear. 
%require clarification
remain unclear.

%%%%
\begin{figure}[tb]
\includegraphics[width=3.375in]{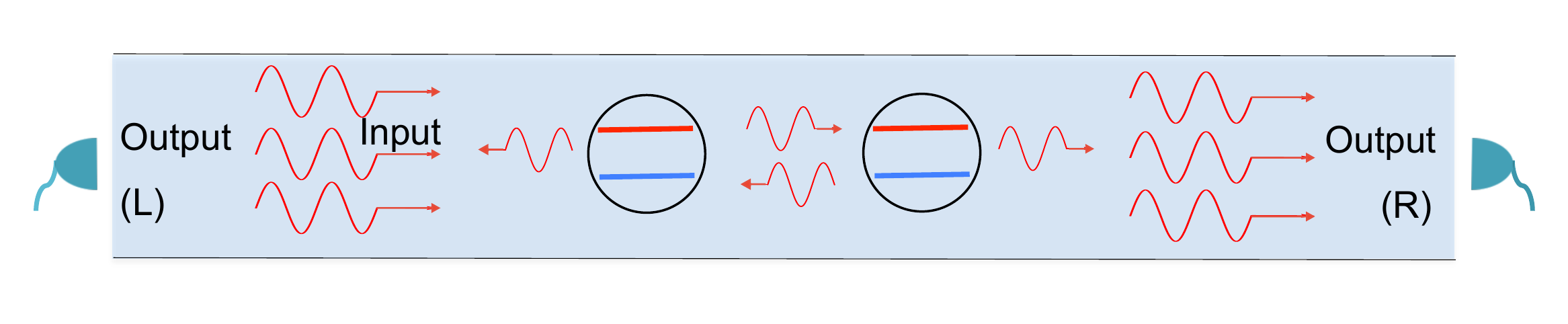}
\caption{Schematic of qubit(s) coupled to a 1D waveguide. A right-going photonic state is input from the left end. After its interaction with the qubit(s), right-going and left-going output states are detected at the right and left ends, respectively.}\label{WQED1Q}
\end{figure}
%%%%

In this work, we study photon statistics in a waveguide interacting with one or two qubits (see Fig.\,\ref{WQED1Q}). We show explicitly that the strong interference between photons confined in a waveguide plays a vital role in quantum jumps and thus in the photon statistics. We find that the statistics of the time interval between two adjacent photons, known as the waiting time, yields a picture that is much clearer than that from the intensity correlation function, providing new insights. In the two qubit case, for instance, a mixture of bunching and antibunching exists. 

The article is organized as follows. First, it is worth noting that jump operators must be chosen carefully in order to faithfully describe photon detections at the outputs instead of simply qubit emission events. Thus we show how to derive jump operators from input-output relations 
%and then use them to elucidate the role of photon interference 
in Sec.\,\ref{Sec:JumpO}. Then in Sec.\,\ref{Sec:InferQJ}, we offer a new view and explanation of the well-known perfect reflection of a single photon off a qubit in waveguide QED in terms of quantum interference and quantum jumps. In Sec.\,\ref{Sec:UpJump}, an explanation of photon bunching is given in terms of interference and a resultant photon absorption process.
%, in contrast to commonly believed stimulated emission process. 
%\HUB{[we have to talk about this again...]} 
Then from the statistics of photon arrival times, quantified by the probability distribution of waiting times and joint probability of adjacent waiting times, we show that photon bunching and antibunching can be clearly and precisely defined, in the case of both a single qubit (Sec.\,\ref{Sec:Statistics}) and two qubits (\ref{Sec:TwoQubits}). A stochastic understanding of $g^{(2)}(\tau)$ through quantum jumps and its relation to the waiting time distribution are given in Sec.\,\ref{Sec:Statistics}. Moreover, in Sec.\,\ref{Sec:TwoQubits}, we observe more complex correlations among the photons, such as the coexistence of bunching and antibunching. This may be useful for the engineering of quantum states of light and the design of photonic quantum gates.

%%%%%%%%% ---------------Input output Relations---------------------- %%%%%%%%%%%%%%%%%%%%%%%%%
\section{Jump Operators from Input-output Relations}
\label{Sec:JumpO}
The system we study consists of qubit(s) coupled to a 1D waveguide under continuous monitoring at the left and right end as shown in Fig.\,\ref{WQED1Q}. After applying the rotating wave approximation, the system is described by the Hamiltonian \cite{FanPRA2010}
\begin{equation}
\begin{split}
H = &\frac{1}{2} \sum_{j} \omega^{(j)}_{\text{eg}} \sigma^{z}_{j} + \int d \omega \,\, \omega \big( r_{\omega}^{\dagger} r_{\omega} - l_{\omega}^{\dagger} l_{\omega} \big)  \\
       &+ \int d \omega \,\,  \sum_{j} g_{j} \big[ e^{i \omega t_{j}} \sigma^{+}_{j} (r_{\omega} + l_{\omega}) + h.c.  \big],
\end{split}
\end{equation}
where $\omega^{(j)}_{\text{eg}}$ and $t_{j}$ are the frequency and position of the $j$-th qubit,  $\sigma^{+}_{j}$ is its raising operator, $r_{\omega} (l_{\omega})$ are the annihilation operator for right(left)-going modes with frequency $\omega$, and $g_{j}$ is the coupling strength between the $j$-th qubit and the waveguide. The decay rate $\Gamma_{j}$ of the $j$-th qubit is then given by $\Gamma_{j} = 4 \pi g_{j}^2$. We start with the single qubit case and then turn to two qubits. 

The input-output relations between the 
%right(left)-going 
output operators, $r_{\text{out}}$ and $l_{\text{out}}$, and the 
%right(left)-going 
corresponding input operators $r_{\text{in}}$ and $l_{\text{in}}$ follow from input-output theory 
%\cite{GardinerPRA1985, FanPRA2010, KocabasPRA2012, LalumierePRA2013}:
\cite{GardinerBook2004, FanPRA2010, LalumierePRA2013}:
\begin{equation}
\begin{split}\label{IO}
r_{\text{out}}(t) &= r_{\text{in}}(t) - i \sqrt{2 \pi} g \sigma^{-} (t), \\
l_{\text{out}}(t) &= l_{\text{in}}(t) - i \sqrt{2 \pi} g \sigma^{-} (t).
\end{split}
\end{equation}
For the input state, we use a right-going monochromatic coherent state  $\ket{\alpha_{k}}$ with frequency $k$, 
%which is defined as 
\begin{equation} \label{coherent}
\ket{\alpha_{k}} = e^{-\frac{1}{2} |\alpha_{k}|^2}  \sum_{n=0}^{\infty} \frac{\alpha_{k}^n}{\sqrt{n !}} \ket{n}.
\end{equation}
It then follows that $r_{\text{in}}(t) \ket{\alpha_{k}} = ( \alpha_{k}/  \sqrt{2 \pi} ) e^{-i k t} \ket{\alpha_{k}}$ and $l_{\text{in}}(t) \ket{\alpha_{k}} = 0 $. The input photon flux $\bar{n}$ is given by $ \bar{n} = |\alpha_{k}|^2/2 \pi$.

In order for our jump operators to correspond to individual outgoing photons, we define them according to Eq.\,\eqref{IO} as $J_{\text{R}}^{-} (t) = i\,r_{\text{out}} (t)$ and $J_{\text{L}}^{-} (t)=i\,l_{\text{out}} (t)$, where we have chosen the overall phase factor to be $i$. More explicitly (in the Schr\"odinger picture),
\begin{equation}\label{JumpO}
J_{\text{R}}^{-} = \sqrt{\Gamma/2}\; \sigma^{-} + i \frac{\alpha_{k}}{\sqrt{2 \pi}} \quad\textrm{and}\quad
J_{\text{L}}^{-} =  \sqrt{\Gamma/2}\; \sigma^{-},
%\begin{split}
%J_{\text{R}}^{-}& = \sqrt{\frac{\Gamma}{2}} \sigma^{-} + i \frac{\alpha_{k}}{\sqrt{2 \pi}},\\
%J_{\text{L}}^{-}& =  \sqrt{\frac{\Gamma}{2}} \sigma^{-},
%\end{split}
\end{equation}
where $\Gamma/2$ is the qubit decay rate to the left or right and $\alpha_{k}/\sqrt{2 \pi}$ is the amplitude of the input coherent state. Note that input operators become complex numbers since our input state is a coherent state. 

The transmitted light described by $J_{\text{R}}^{-}$ is clearly a coherent superposition of photons emitted by the qubit and photons from the input state. As mentioned in the introduction, jump operators that include photon interference have been used to describe the superposition of an output field and a coherent field in homodyne detection \cite{CarmichaelBook2008, WisemanBook2014}, the superposition of fields in cascaded systems \cite{CarmichaelPRL1993, PlenioRMP1998, GardinerBook2004, MirzaJOSAB2013}, quantum trajectories of propagating Fock states \cite{BaragiolaPRA2017}, and, recently, photon detection at the output ends of a waveguide system \cite{ManzoniNatComm2017}. Generally, we thus conclude that input-output relations can be used as a systematic way to find the correct jump operators to describe interference between quantum objects of complex systems such as in circuit QED \cite{SchoelkopfNature2008} or quantum networks \cite{KimbleNature2008, CombesAPX2017}.

%%%%%%%%----------------Quantum Jumps---------------------------%%%%%%%%%%%%%%%%%%%%%%%%%%
\section{Quantum Jumps and Quantum Interference for One Qubit}
\label{Sec:QJ}

%With the jump operators  $J_{\text{R}}^{-}$ and $J_{\text{L}}^{-}$ defined above, we write a master equation of Lindblad form (for details see Appendix \ref{Appenddix:ME}):
%\begin{equation}\label{ME}
%\frac{\partial}{\partial t} \rho_{\text{s}} = - i \Big[ \frac{\Delta}{2} \sigma^{z} + \frac{g}{2} (\alpha_{k} \sigma^{+} +  \alpha_{k}^{\star} \sigma^{-}) , \rho_{\text{s}} \Big] + \mathcal{L}_{\text{R}} \rho_{\text{s}} + \mathcal{L}_{\text{L}} \rho_{\text{s}},
%\end{equation}
%where $\Delta \equiv \omega_{\text{eg}} - k$ is the detuning and $\mathcal{L}_{\text{R(L)}}$ is the Lindblad superoperator given by
%\begin{equation}\label{ME1}
%\mathcal{L}_{\text{R(L)}} \rho_{\text{s}} =  J_{\text{R(L)}}^{-} \rho_{\text{s}} J_{\text{R(L)}}^{+} - \frac{1}{2} \{ J_{\text{R(L)}}^{+} J_{\text{R(L)}}^{-} , \rho_{\text{s}} \} ,
%\end{equation}
%which describes the detection of right(left)-going photons. The simulation then follows the standard quantum jump technique \cite{PlenioRMP1998} 
%\HUB{[add Carmichael ref?]}
%with two channels described by the jump operators $J_{\text{R}}^{-}$ and $J_{\text{L}}^{-}$ (for details see Appendix \ref{Appenddix:QJ}).

With the jump operators  $J_{\text{R}}^{-}$ and $J_{\text{L}}^{-}$ defined above, we write a master equation of Lindblad form. The Lindblad superoperator that describes the detection of right-going photons is \begin{equation}\label{ME1}
\mathcal{L}_{\text{R}} \rho_{\text{s}} =  J_{\text{R}}^{-} \rho_{\text{s}} J_{\text{R}}^{+} - \frac{1}{2} \{ J_{\text{R}}^{+} J_{\text{R}}^{-} , \rho_{\text{s}} \} ,
\end{equation}
and an analogous expression holds for left-going jumps.
% with $\text{R}$ replaced by $\text{L}$. 
The master equation in a rotating frame is (for details see Appendix \ref{Appenddix:ME}):
\begin{equation}\label{ME}
\frac{\partial}{\partial t} \rho_{\text{s}} = - i \Big[ \frac{\Delta}{2} \sigma^{z} + \frac{g}{2} (\alpha_{k} \sigma^{+} +  \alpha_{k}^{\star} \sigma^{-}) , \rho_{\text{s}} \Big] + \mathcal{L}_{\text{R}} \rho_{\text{s}} + \mathcal{L}_{\text{L}} \rho_{\text{s}},
\end{equation}
where $\Delta \equiv \omega_{\text{eg}} - k$ is the detuning. The simulation then follows the standard quantum jump technique \cite{PlenioRMP1998, CarmichaelBook2008,WisemanBook2014} with two channels described by the jump operators $J_{\text{R}}^{-}$ and $J_{\text{L}}^{-}$; see Appendix \ref{Appenddix:QJ} for details.

\begin{figure}
\includegraphics[scale=0.6]{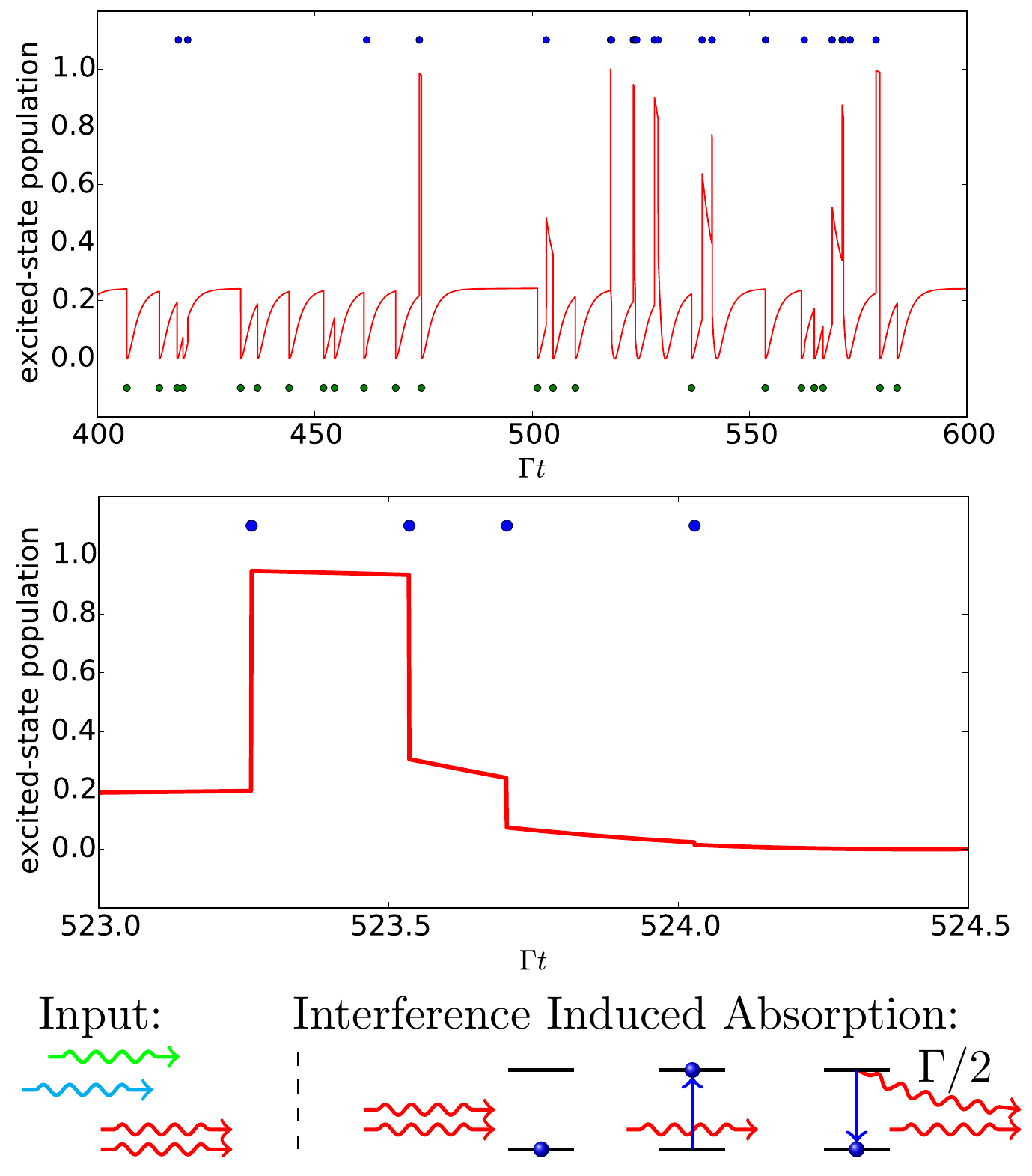}%
\llap{
  \parbox[b]{5.3in}{(a)\\\rule{0ex}{2.9in}
  }}
\llap{
  \parbox[b]{5.3in}{(b)\\\rule{0ex}{1.4in}
  }}  
\llap{
  \parbox[b]{5.2in}{(c)\\\rule{0ex}{0.5in}
  }}  
\caption{(a) One trajectory given by the quantum jump simulation for one qubit. Blue dots at the top mark detections of right-going (transmitted) photons; green dots at the bottom mark those for left-going (reflected) photon. The red line is the excited-state population of the qubit. (b) A zoomed-in view of a cluster of photons and successive jumps. Note in particular the jump \emph{up}. (c) A schematic of the up jump process. For a weak coherent state, which is dominated by one and two-photon components, a jump up happens when one photon is transmitted (detected) and the qubit wavefunction collapses to the excited state, which implies that a photon has been absorbed. Subsequently, the spontaneous decay of the qubit to the right makes it possible to detect two photons at the same time, which is the photon bunching phenomenon.
(Parameters: $\alpha=1$, $\Delta =0$.)}\label{Trajectory}
\end{figure}

A single trajectory given by the quantum jumps simulation is shown in Fig.\,\ref{Trajectory}. Fig.\,\ref{Trajectory}(a) shows the time evolution of the excited-state population: there are jumps corresponding to abrupt changes. At each jump one photon is detected at either the right or the left end. Note the interesting feature that there are not only jumps down as expected from usual quantum jump simulations but also \emph{jumps up}---a sudden \emph{increase} of excited-state population \cite{CarmichaelPRL1993}. A zoomed-in view of a cluster of photons is shown in Fig.\,\ref{Trajectory}(b): clearly, there can be multiple successive jumps down following directly after a jump up. The up jumps are a signature of the back-action of monitoring at the right end. In this back-action, the interference between the photons emitted from the qubit and those from the input is encoded, as can be seen from the expression for $J_{\text{R}}^{-}$ in Eq.\,\eqref{JumpO}. 

%%%%
%%%%

\subsection{Quantum Interference Revealed by Quantum Jumps}
\label{Sec:InferQJ}

These unusual features of the quantum jump trajectories can be understood by considering the wavefunction of the qubit before and after a jump. 
The continuous part of a trajectory is governed by a non-Hermitian effective Hamiltonian
\begin{equation}
H_{\text{eff}} = \frac{1}{2} \Delta \sigma^{z} + g \alpha_{k} \sigma^{+} - i \frac{1}{2} \Gamma \sigma^{+} \sigma^{-} - i \frac{1}{2} \bar{n},
\end{equation}
where there are two imaginary contributions describing the detection of, first, photons emitted by the qubit with rate $\Gamma$ and, second, photons from the input coherent state with rate $\bar{n}$. 
%{\color{cyan} There are $1/2$ factors for them because $H_{\text{eff}}$ describes the amplitude of these processes.} 
$\Delta=0$ under resonant driving. In the absence of jumps in the interval $(t_{0}, t)$, the time evolution of the unnormalized wavefunction $\ket{\tilde{\psi} (t)} $ is given by 
%the Schr\"odinger equation 
$\ket{\tilde{\psi} (t)} = \exp[ - i H_{\text{eff}} (t - t_{0}) ]\ket{\tilde{\psi} (t_{0})}$. The normalized wavefunction is then $ \ket{\psi (t)}= c_e (t) \ket{e} + c_g (t) \ket{g}$ with
\begin{equation}
\frac{c_e (t)}{c_g (t)} = \frac{c_e (t_0)}{c_g (t_0)} e^{- \frac{\Gamma}{2} (t-t_0)} - i \sqrt{\frac{2}{\Gamma}} \frac{\alpha_{k}}{\sqrt{2 \pi}}   \big(1 - e^{-\frac{\Gamma}{2} (t-t_0) } \big) .
\end{equation}
The corresponding probability density of a right jump is given by $p_{\text{R}}=\braket{\psi | J^{+}_{\text{R}} J^{-}_{\text{R}} | \psi }$ and can be shown to be
\begin{equation} \label{PR}
p_{\text{R}} = \bar{n} + \frac{\Gamma}{2} |c_e|^2 + \bigg(i \frac{\alpha_{k}}{\sqrt{2\pi}} \sqrt{\frac{\Gamma}{2}}  c_e^{*} c_g + \mathrm{c.c.} \bigg),
\end{equation}
where the third term is a cross term due to photon interference. In the limit of weak driving, 
%$c_g \approx 1$ and $c_e \approx  - i  \sqrt{2/\Gamma} ( \alpha_{k} / \sqrt{2 \pi}) c_g$ 
\begin{equation}
c_g \approx 1 \quad\text{and}\quad c_e \approx  - i  \sqrt{2/\Gamma} ( \alpha_{k} / \sqrt{2 \pi}) c_g 
\label{eq:cgceweak}
\end{equation}
as the exponential factors vanish quickly. In that case, $p_{\text{R}} \approx 0$ since the cross term cancels the first two terms in \eqref{PR} due to the the $-i$ factor between $c_e$ and $c_g$. This destructive interference between emitted photons and input photons explains, in the quantum trajectory description, the well-known low on-resonance transmission rate under weak driving \cite{Chang2007NatPhys, AstafievScience2010, ZhengPRA10, HoiPRL11}. When off-resonance, $\Delta \neq 0$, the phase difference is time dependent and so the destructive interference relation is broken, thus leading to increased transmission.

\subsection{Up Jumps and Interference Induced Absorption}
\label{Sec:UpJump}

After a right jump, the wavefunction collapses to $\ket{\psi'} = J^{-}_{\text{R}} \ket{\psi} / \big\| J^{-}_{\text{R}} \ket{\psi}  \big\| = c'_e \ket{e} + c'_{g} \ket{g}$.  
One finds that
\begin{equation}
\abs{ \frac{c_{\text{e}}'}{c_{\text{g}}'} } = \Bigg| \frac{ 1 }{ c_g /c_e   - i \frac{\sqrt{ \Gamma\pi }}{\alpha_{k} }  } \Bigg|,
\end{equation}
whose denominator vanishes in the weak driving limit based on \eqref{eq:cgceweak}. That is, \emph{the qubit jumps up to the excited state}. 

Once it is in the excited state $\ket{e}$, it is very likely to have a second jump, either to the left or right. If it is a right jump, the wavefunction collapses to $ i \alpha_{k}/\sqrt{2\pi} \ket{e} +  \sqrt{\Gamma/2} \ket{g}$---note that the phase between the two terms has flipped from $-i$ to $i$ and so the interference in $p_R$ in \eqref{PR} becomes \emph{constructive}. Therefore, it is possible to have a third jump to the right. \emph{This phenomenon of photon bundles heralded by an up jump, caused by photon interference, 
%leads to the bunching effect of photons.} 
is the mechanism for bunching of the transmitted photons.} 

A schematic of the up jump process and the resultant photon bunching is shown in Fig.\,\ref{Trajectory}(c) for an input coherent state (first panel). 
The continuous evolution generated by $H_\text{eff}$ in the two photon sector produces a state that is a superposition of the ground and excited states, though predominantly ground state for a weak coherent input (second panel). The jump operator $J^-_\text{R}$ corresponds to the presence of a transmitted photon in the waveguide at the location of the qubit; furthermore, $J^-_\text{R}$ collapses the wavefunction onto the excited-state portion of the wavefunction, thus producing the sudden increase of excited-state population that accompanies photon detection (third panel). This is reasonable since transmission can only occur when the qubit is in its excited state (a single on-resonant photon is perfectly reflected from the ground state). Finally, the qubit decays to the right at the rate $\Gamma/2$ (fourth panel). As discussed in the last paragraph, this latter process is again heavily influenced by interference effects. 

\begin{figure}
\includegraphics[scale=0.35]{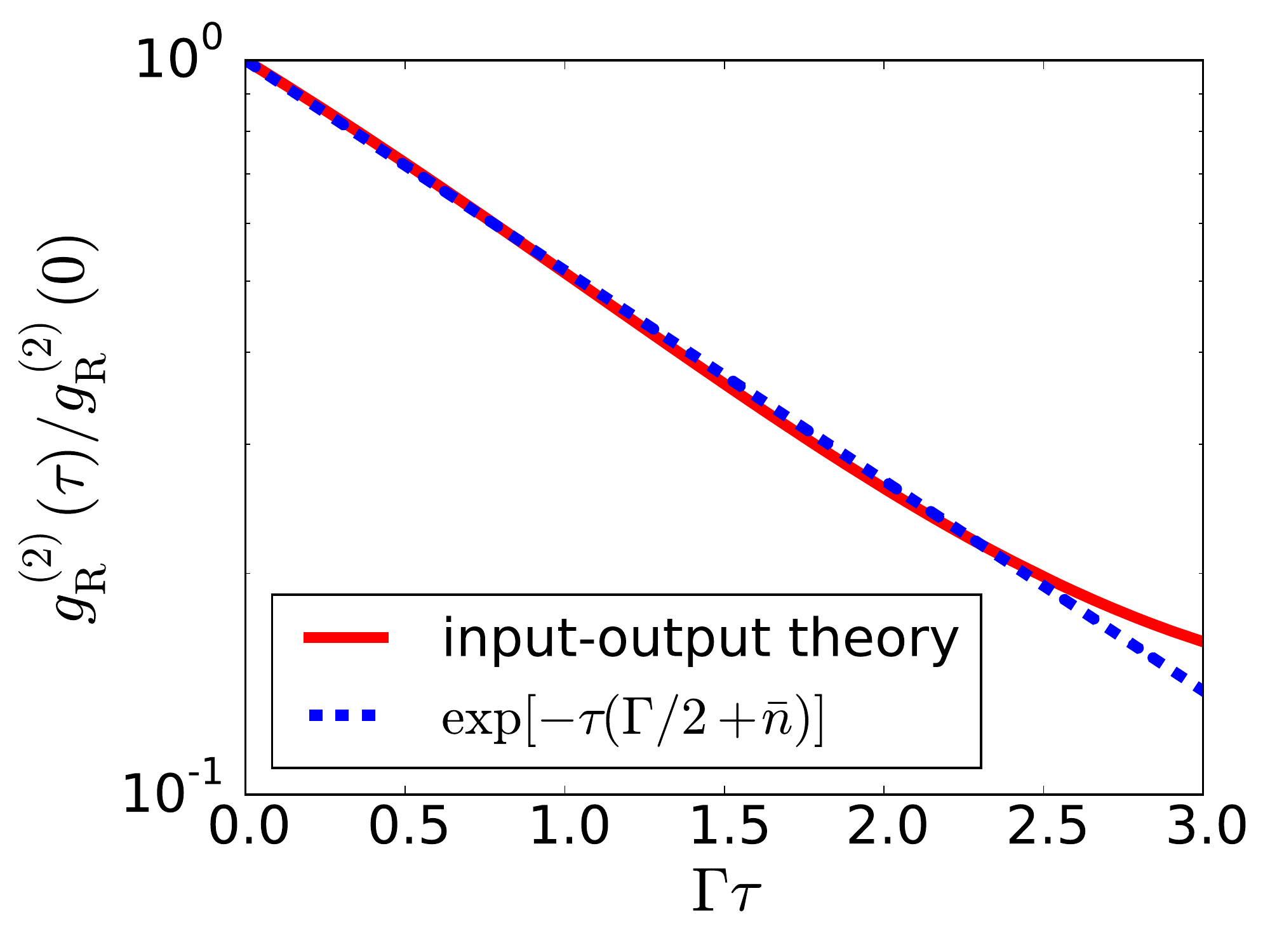}
\caption{Comparison between the second order correlation function $g^{(2)}_{\text{R}}(\tau)$ for one qubit calculated from input-output theory and the exponential relation $g^{(2)}_{\text{R}}(\tau) = g^{(2)}_{\text{R}}(0) \exp[-\tau(\Gamma/2+\bar{n})]$. They agree with each other when $\tau$ is not very large, which verifies that the second photon of a bunched photon pair comes from the spontaneous decay of the excited state and direct transmission of the input coherent state.
 (Parameters: $\alpha=1$, $\Delta =0$.)} 
\label{expg2}
\end{figure}

Given the detection of the transmitted photon at time $t$, the probability density of detecting another photon at time $t+\tau$ is given by $P(\text{jump at } t+\tau | \text{ jump at } t) = (\Gamma/2+\bar{n}) \exp[-\tau(\Gamma/2+\bar{n})]$, where $\Gamma/2$ corresponds to the spontaneous decay of an exited state and $\bar{n}$ corresponds to the direct transmission of the input coherent state as the qubit is excited. Since the second-order correlation function $g^{(2)}(\tau)$ equals this conditional probability normalized by the very small photon detection probability of transmitted light, as discussed further in the next section, $g^{(2)}(0)$ will be very large and $g^{(2)}(\tau)$ then decays exponentially under weak driving i.e.\, $g^{(2)}(\tau) = g^{(2)}(0) \exp[-\tau(\Gamma/2+\bar{n})]$. A comparison between this exponential relation and $g^{(2)}(\tau)$ calculated from input-output theory is shown in Fig.\,\ref{expg2} and they agree with each other when $\tau$ is not very large. Thus, in this context of photon detection at precise times, it is simply the {\it spontaneous decay} of the excited qubit and the direct transmission of the input state that gives the probability to detect two photons together.

\section{Photon Statistics and Photon Correlations}
\label{Sec:Statistics}

Since the full time series of photon detection events is available from our quantum jump simulation, any desired photon counting statistic can be calculated. Here we focus on two: first, the waiting time distribution (WTD) $\mathcal{W} (\tau)$ \cite{PlenioRMP1998, CarmichaelBook2008}, which is the probability distribution of the time interval between two successive photon arrivals (i.e.\,jumps), and, second, the adjacent waiting time distribution (AWTD) $\mathcal{A}(\tau_{1},\tau_{2})$, which is the joint probability density of the two adjacent waiting times $\tau_{1}$ and $\tau_{2}$.

%%%%
\begin{figure}
\includegraphics[scale=0.60]{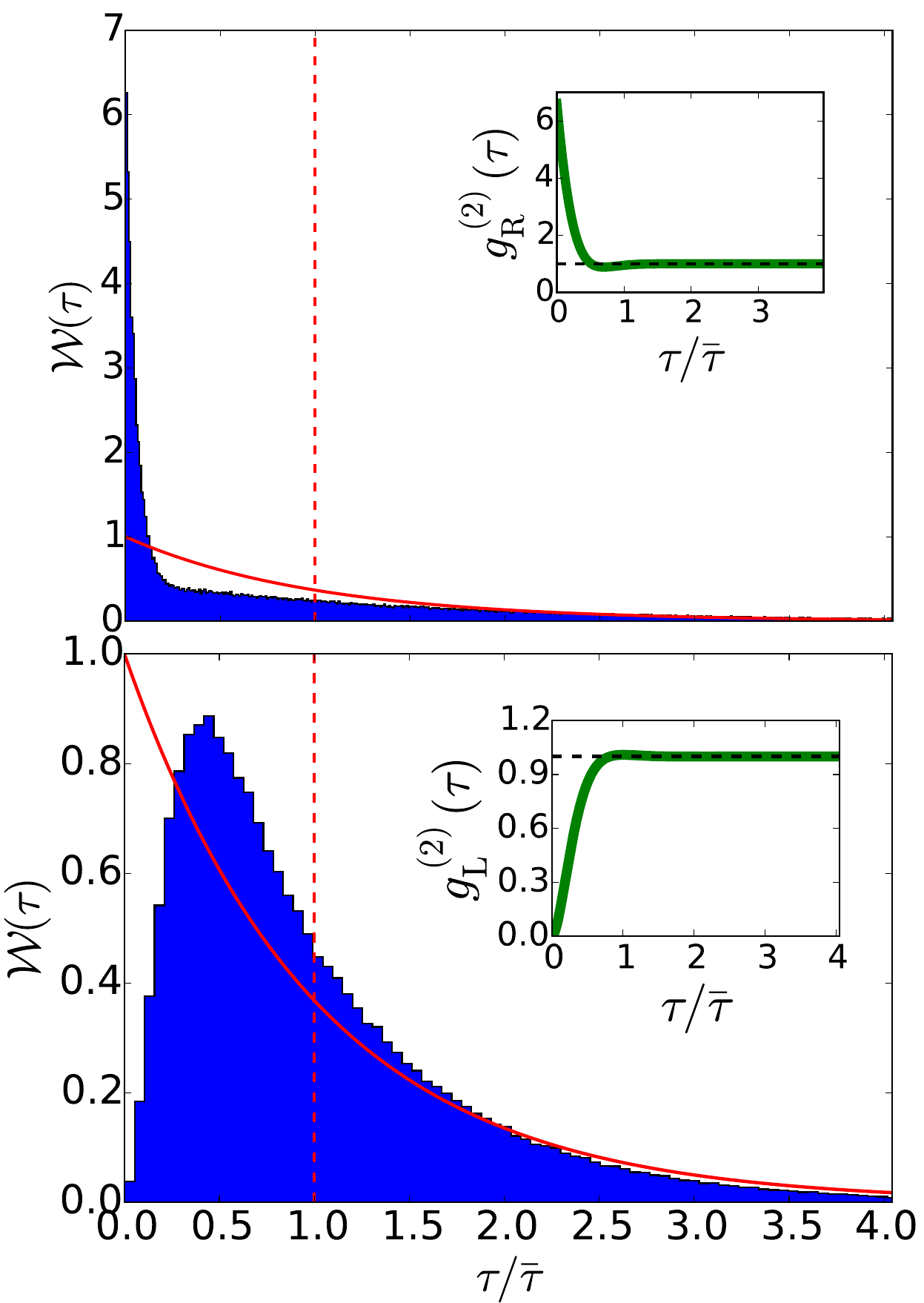}
\llap{
  \parbox[b]{2.0in}{(a): transmitted\\\rule{0ex}{2.8in}
  }}
\llap{
  \parbox[b]{2.0in}{(b): reflected\\\rule{0ex}{0.7in}
  }}  
\caption{Waiting time distribution (WTD) and $g^{(2)}(\tau)$ (insets) for (a) transmitted (right-going) and (b) reflected (left-going) photons for one qubit. Time is scaled by their respective average waiting time $\bar{\tau}$ (marked by red dashed line). The red solid line is the WTD for the input coherent state, which is Possionian. The large peak near $0$ for transmission indicates strong bunching, while the dip for reflection indicates anti-bunching. (Parameters: $\alpha=1$, $\Delta =0$.)}
\label{1QWTD}
\end{figure}

%%%%

The WTD for transmitted (right-going) and reflected (left-going) photons are shown in Fig.\,\ref{1QWTD}. Time is normalized to the mean waiting time $\bar{\tau}$ as the mean transmission or reflection is not relevant for photon bunching or antibunching. Clearly, the statistics of photons after their interaction with the qubit is very different from the input coherent state. For transmitted photons, there is a large peak at $\tau=0$, which means that photons are more likely to arrive at the same time, i.e.\,they are bunched. For reflected photons, the WTD peaks around a value between $0$ and $\bar{\tau}$, implying that photons tend to arrive with some separation, i.e.\,they are anti-bunched. This latter property is simply because, as is well known, every left jump necessarily takes the wavefunction to the ground state from which some time is required to again become excited. 

The second-order correlation function $g^{(2)}(\tau)$ calculated from  input-output theory is also shown in Fig.\,\ref{1QWTD} (see insets). The general conclusion from the WTD is born out by this statistic as well: 
for transmitted light, $g^{(2)}(\tau)$ decreases from a value larger than $1$ to $1$ while for reflected light it increases from $0$ to $1$.
This agreement is not surprising if we look at $g^{(2)}(\tau)$ in terms of quantum jumps. From the input-output relation and definition of jump operators,  Eqs.\,\eqref{IO} and \eqref{JumpO}, $g^{(2)}(\tau)$ can be written as
\begin{equation}
g^{(2)}_{i} (\tau) = \frac{\braket{J_{i}^{+}(t) J_{i}^{+}(t+\tau) J_{i}^{-}(t+\tau) J_{i}^{-}(t)}}{\braket{J_{i}^{+}(t) J_{i}^{-}(t)} \braket{ J_{i}^{+}(t+\tau) J_{i}^{-}(t+\tau) } },                         
\end{equation}                             
where $i=\text{R}, \text{L}$ for transmitted and reflected photons respectively. After transforming from the Heisenberg to the Schr\"odinger picture, $ J_{i}^{-}  \rho_{\text{s}}(t) J_{i}^{+} / \text{Tr}\{ J_{i}^{-}  \rho_{\text{s}}(t) J_{i}^{+} \}  $ is the normalized density matrix after a jump at time $t$. Denoting the time evolution given by the master equation \eqref{ME} by $\mathcal{M}_{\tau}$, $\rho_{\text{s}} (t+\tau) = \mathcal{M}_{\tau}[ \rho_{\text{s}} (t) ]$, we get
\begin{equation}
g^{(2)}_{i} (\tau) =\frac{ \text{Tr}\{ J_{i}^{-} ( \mathcal{M}_{\tau}[ \frac{J_{i}^{-}  \rho_{\text{s}}(t) J_{i}^{+}}{\text{Tr}\{ J_{i}^{-}  \rho_{\text{s}}(t) J_{i}^{+} \} } ] ) J_{i}^{+}  \}  }{ \text{Tr}\{ J_{i}^{-} \rho_{\text{s}}(t+\tau) J_{i}^{+} \} }.
\end{equation}
The numerator describes the probability density of detecting a photon at time $t+\tau$ conditioned on a detection at time $t$, while the denominator describes the probability density of detecting a photon at time $t+\tau$ without any prior condition. 
%In the language of probability,
%\begin{equation}
%g^{(2)}_{i} (\tau) = \frac{ P(\text{jump at } t+\tau | \text{ jump at } t) }{ P(\text{jump at } t+\tau) }.
%\end{equation}
When the photon flux is normalized to be $1$, the denominator is $1$, yielding in the language of probability $g^{(2)}_{i} (\tau) = P(\text{jump at } t+\tau | \text{ jump at } t) $. $g^{(2)}(\tau)$ can be computed in this statistical sense and is shown in Appendix \ref{Appenddix:g2}.

For the WTD $\mathcal{W}(\tau)$, however, there is an additional ``next photon'' requirement that there be no photon detected between $t$ and $t+\tau$ \cite{CarmichaelBook2003, EmaryPRB2012}. 
Because of this extra condition, $\mathcal{W}(\tau) \le g^{(2)}(\tau)$. In fact, the bound is reached as $\tau \rightarrow 0$,  $g^{(2)}(0) = \mathcal{W}(0)$, as there can be no intervening photon  
\footnote{Note that $\mathcal{W}(\tau)$ here is the probability density function (PDF) of the variable $\tau / \bar{\tau} $ after the normalization of time. Denote the PDF of the variable $\tau$ by $W(\tau)$. Then $\mathcal{W}(\tau)$ = $\bar{\tau} W(\tau)$. That is, $g^{(2)}(0) = \bar{\tau} W(0)$.}. 
With regard to bunching and anti-bunching, since 
$g^{(2)}(0)>1$ $(<1)$ tells us that it is more (less) likely to detect another photon immediately after a detection, in some events photons are bunched (antibunched). However, in $g^{(2)}(\tau)$ the detailed information about the distribution of bunching and anti-bunching contained in the WTD is missing. Indeed, in order to determine the WTD, all orders of correlation functions are needed \cite{CarmichaelPRA1989}. When photon statistics is simple, as in the case of a single qubit, the qualitative pictures from $\mathcal{W}(\tau)$ and $g^{(2)}(\tau)$ agree. But, as we now show, for complex photon statistics there will be differences.
%footnote explains PDF of tau

%%%%%%%%%%---------------Two Qubits----------------------------------%%%%%%%%%%%%%%%%%%%%%%%%
\section{Coexistence of Bunching and Antibunching for Two Qubits}
\label{Sec:TwoQubits}

To study more complex photon statistics, we carry out quantum jump simulations for two identical qubits with a small separation $\Delta t$. The cooperative effect of two qubits in a 1D waveguide was studied previously in, e.g., Refs.\,\cite{LalumierePRA2013, ZhengPRL13, LaaksoPRL14, RoyRMP2016, GuPhysRep17} but not from the quantum jump viewpoint. From the input-output relation for two qubits (see Appendix \ref{Appenddix:ME}), we define the jump operators as
\begin{equation}\label{2QJO}
\begin{split}
J_{\text{R}}^{-}& =\sqrt{ \frac{\Gamma}{2} } \sigma^{-}_{1} e^{i \omega_{\text{eg}} \Delta t} + \sqrt{ \frac{\Gamma}{2} } \sigma^{-}_{2} +  i \frac{\alpha_{k}}{\sqrt{2 \pi}} e^{i k \Delta t}, \\
J_{\text{L}}^{-}& =\sqrt{ \frac{\Gamma}{2} } \sigma^{-}_{1} + \sqrt{ \frac{\Gamma}{2} }  \sigma^{-}_{2} e^{i \omega_{\text{eg}} \Delta t }.
\end{split}
\end{equation}
The phases $\omega_{\text{eg}} \Delta t$ and $k \Delta t$ are due to the time delay of photons traveling from one qubit to another. The Markovian approximation \cite{LalumierePRA2013} has been applied to deal with this time delay in that only the constant frequency $\omega_{eg}$ appears in  the qubit phase factors. In these jump operators, in addition to interference between input photons and photons emitted from qubits, there is clearly also interference between photons emitted from the two qubits.

%%%%
%\begin{figure}
%\includegraphics[scale=0.65]{Fig5WTDg2.pdf}
%\caption{Photon statistics for two qubits with separation $1/4$ wavelength for reflected (left-going) photons. (a) Waiting time distribution (WTD) and $g^{(2)}(\tau)$ (inset); (b) adjacent waiting time distribution (AWTD). Note the clear coexistence of bunching and anti-bunching. (Parameters: $\alpha=1$, $\Delta =0$, $k\Delta t = \pi/2$.)  }\label{WTDg2}
%\end{figure}
\begin{figure}
\includegraphics[scale=0.65]{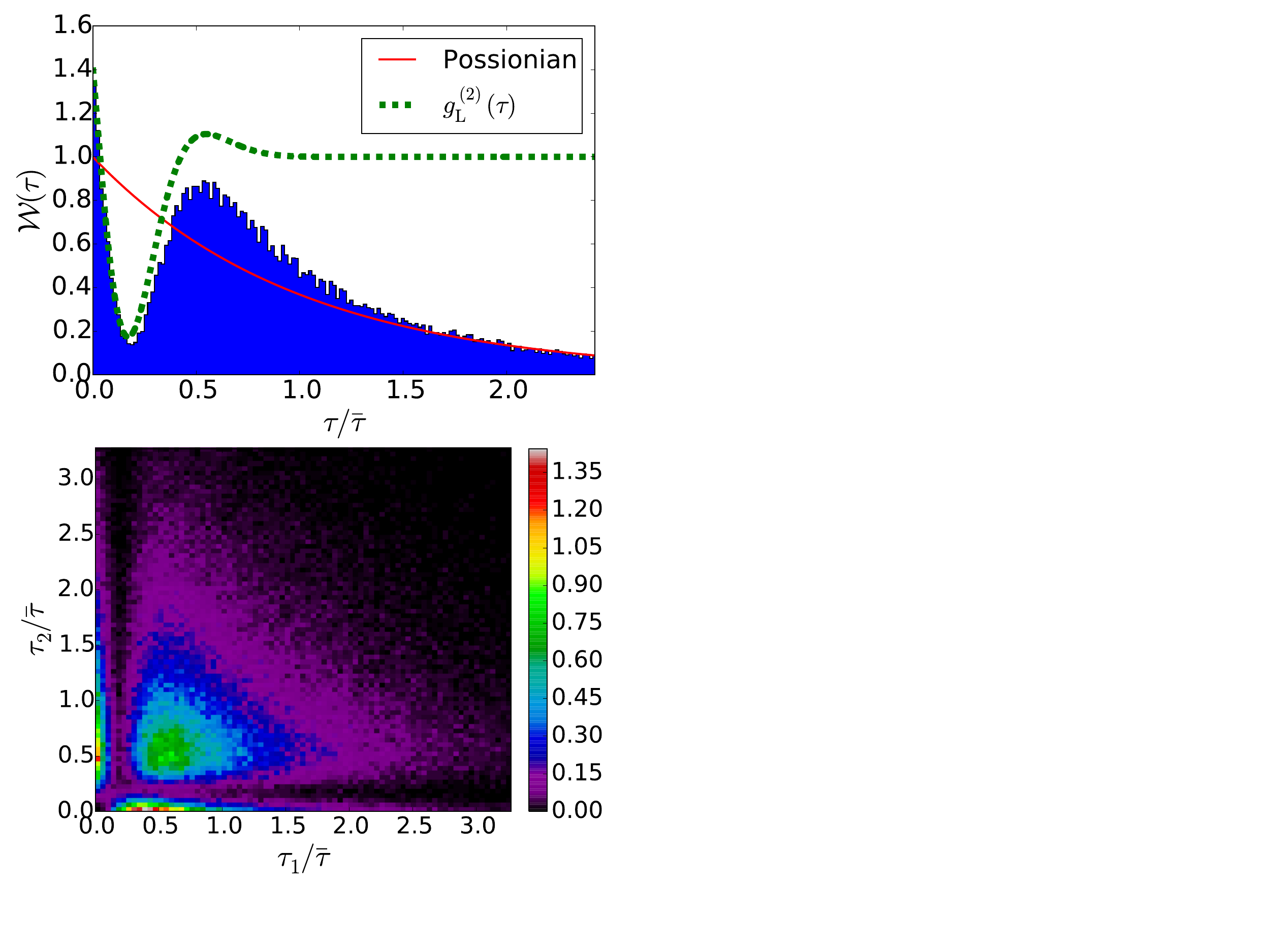}
\llap{
  \parbox[b]{4.8in}{(a)\\\rule{0ex}{4.0in}
  }}
\llap{
  \parbox[b]{1.6in}{\color{white}{(b)}\\\rule{0ex}{1.9in}
  }}  
\caption{Photon statistics for two qubits with separation $1/4$ wavelength for reflected (left-going) photons. (a) Waiting time distribution (WTD) and $g^{(2)}(\tau)$; (b) adjacent waiting time distribution (AWTD). Note the clear coexistence of bunching and anti-bunching. (Parameters: $\alpha=1$, $\Delta =0$, $k\Delta t = \pi/2$.)  }\label{WTDg2}
\end{figure}
%%%%

We study the statistics of reflected photons for two qubits with separation $1/4$ wavelength under resonant driving, that is, $k=\omega_{\text{eg}}$ and $k \Delta t = \pi/2$. The WTD $\mathcal{W}(\tau)$ from quantum jump simulations and $g^{(2)}(\tau)$ calculated by input-output theory are shown in Fig.\,\ref{WTDg2}(a). The WTD has \emph{two} peaks, one at $0$ and the other at a value between $0$ and $1$. \emph{It is thus a combination of the bunching and antibunching statistics shown in Fig.\,\ref{1QWTD} and cannot be simply categorized as either bunching or antibunching alone.} However, $g^{(2)}(\tau)$ is larger than 1 at $\tau=0$ and then decreases, which means that the photon statistics is bunching according to some definitions \cite{CarmichaelBook2003}. This is a physical example showing that the categorization of photon statistics into bunching or antibunching according to $g^{(2)}(\tau)$ is 
inaccurate.
%not precise. 

To investigate further these complex photon statistics, we present the adjacent waiting time distribution (AWTD) $\mathcal{A}(\tau_{1},\tau_{2})$---the joint probability density of two adjacent waiting times $\tau_{1}$ and $\tau_{2}$---in Fig.\,\ref{WTDg2}. Note that $\mathcal{A}(0,0) =0$ since 
%it is impossible to detect three photons at the same time for light reflected by two qubits. 
two qubits can reflect at most two photons at the same time.
$\mathcal{A}(\tau_{1},\tau_{2})$ has three islands of significant weight: one at the center and two along the axes. Therefore for three photons it is more likely to have two of them anti-bunched while the third can be either bunched or anti-bunched to them. Clearly, the photon statistics is a mixture of bunching and antibunching characteristics.

Quantum jump processes yield insight into the origin of these statistics. 
For $k \Delta t=\pi/2$, $J^{-}_{\text{L}} \sim (\sigma^{-}_{1} + i  \sigma^{-}_{2})$ leads to jumps $\ket{ee} \rightarrow \ket{+i}$ and $\ket{+i} \rightarrow \ket{gg}$, where $\ket{\pm i} = ( \ket{ge} \pm i \ket{eg} ) /\sqrt{2}$. In a left jump, the wavefunction $c_1 \ket{ee} + c_{2} \ket{+i} + c_3 \ket{-i} + c_4 \ket{gg}$ collapses to $\big( c_1 \ket{+i} + i c_2 \ket{gg} \big)/ \sqrt{| c_1 |^2 + | c_2| ^2}$ with probability density $p_{\text{L}} =\Gamma ( | c_1 |^2 + | c_2| ^2 )$. If $| c_1| \ll | c_2|$, there is no second jump, leading to photon antibunching. However, if $| c_1 | \gg | c_2|$, there is likely to be a second jump, producing photon bunching. And those two situations may have similar jump probability density $p_{\text{L}}$. This explains the two peaks of the WTD shown in Fig.\,\ref{WTDg2}(a).

If there are two jumps at the same time, then the system is in $\ket{gg}$, after which some time is required to have a third jump. That is, if $\tau_1$ is small then $\tau_2$ must be large. But if the system evolves for some time between the first and second jump, i.e.\,$\tau_1$ is large, then it is possible to have $| c_1 |$ larger or smaller than $| c_2|$, which leads to a small and large $\tau_2$ respectively. These three cases explain the three islands in the AWTD shown in Fig.\,\ref{WTDg2}(b).

In $g^{(2)}(\tau)$ there are some veiled signatures of the complex photon statistics, as can be seen in Fig.\,\ref{WTDg2}(a) by comparing it to $\mathcal{W}(\tau)$. After decreasing to a value less than $1$, $g^{(2)}(\tau)$ slightly overshoots $1$ before approaching it asymptotically. Since $g^{(2)}(\tau) \ge \mathcal{W}(\tau)$, the valley in $g^{(2)}(\tau)$ implies a valley in $\mathcal{W}(\tau)$ and so indicates an anti-bunching component to the statistics. 
However, the fact that the most likely separation between photons (other than $0$) is near $\bar{\tau}/2$ is completely missed in $g^{(2)}(\tau)$, a rather dramatic failure to provide a good qualitative picture of the complex photon statistics present. 

%The oscillating behaviors of $g^{(2)}(\tau)$ means that there are multiple peaks in the WTD. From this we claim that the value of $g^{(2)}(\tau)$ is not the only quantity that matters. {\color{red} The shape of it also matters, which could indicate the shape of the WTD.} 

%%%%%%%%------------------Conclusion-------------------------------%%%%%%%%%%%%%%%%%%%%%%%%%%
\section{Conclusion}
To correctly describe the detection of single photons %at the output ends of 
in waveguide QED, we have defined jump operators from input-output relations that correctly account for the interference between different components of photons. From quantum jumps, the meaning of photon bunching and antibunching can be understood more clearly and precisely using the WTD $\mathcal{W}(\tau)$ and AWTD $\mathcal{A}(\tau_{1},\tau_{2})$. We showed, for instance, that bunching in waveguide QED is caused by successive down jumps heralded by an up jump. For 
%more complex system such as 
two qubits in a 1D waveguide, complex photon statistics such as a mixture of bunching and antibunching was discovered, statistics that cannot be simply categorized as bunching or antibunching.
% with $g^{(2)}(\tau)$. 
%For such complex photon statistics, the shape of $g^{(2)}(\tau)$ is important since the shape of it can indicate the peaks and valleys of the WTD. 

We expect that for other complex systems the photon statistics will also be too complex to be precisely and clearly characterized by $g^{(2)}(\tau)$, not to mention $g^{(2)}(0)$. The distribution and correlations of waiting times provide more precise and clearer methods to characterize photon statistics and to go beyond the simple-minded categorization of  bunching and antibunching. Although we used waveguide QED as an example system, our method and conclusions are very general and can be adapted to other optical and electronic systems \cite{EmaryPRB2012, RajabiPRL2013}. For the future, given the current interest in non-Markovian effects \cite{BreuerRMP2016, deVegaRMP2017}, non-Markovian quantum jumps \cite{PiiloPRL2008} offer a possible method to study the relation between photon statistics and non-Markovianity \cite{LuomaPRA2012, ThomasPRB2013}. In addition, previous quantum jump studies of a large number of qubits reveal quantum many-body effects \cite{LeePRL2012, ManzoniNatComm2017}, whose connection with photon statistics can be studied with our method.

\begin{acknowledgments}
We thank Yao-Lung Leo Fang and Kenneth Brown for valuable discussions. This work was supported by U.S.\ NSF Grant No.~PHY-14-04125.
\end{acknowledgments}

\appendix

\section{Derivation of the Master Equations}
\label{Appenddix:ME}

%%%%%%%%------------------Master Equation-------------------------------%%%%%%%%%%%%%%%%%%%%%%%%%%

With the input-output relation shown in Eq.\,\eqref{IO} and the input coherent state defined in Eq.\,\eqref{coherent}, 
%of the main text, 
we redefine $r(t) = r(t) e^{-i k t}$, $l(t) = l(t) e^{-i k t}$ and $\sigma_{-}(t) = \sigma_{-}(t) e^{-i k t}$ in an interaction picture with respect to $\frac{1}{2} k \sigma_{z}$ corresponding to a rotating frame. The input-output relation then becomes
\begin{equation}
\begin{split}
r_{\text{out}} (t) &= \frac{\alpha_{k}}{\sqrt{2 \pi}} - i \sqrt{2 \pi} g \sigma_{-} (t),\\
l_{\text{out}} (t) &= - i \sqrt{2 \pi} g \sigma_{-} (t).
\end{split}
\end{equation}
By including the coherent state driving in the unitary evolution, one finds a  master equation that describes the decay of a qubit in a waveguide \cite{LalumierePRA2013}, 
\begin{equation} \label{SME}
\frac{\partial}{\partial t} \rho_{\text{s}} = - i \left[ \frac{1}{2} \Delta \sigma_{z} + g \alpha_{k} \sigma_{+} + g \alpha_{k}^{\star} \sigma_{-} \, , \,\rho_{\text{s}} \right] + \sum_{i=R,L}\mathcal{\tilde{L}}_{\text{i}} \rho_{\text{s}}, 
\end{equation} 
with the Lindblad superoperators 
$\mathcal{\tilde{L}}_\text{R}$ and $\mathcal{\tilde{L}}_\text{R}$ that 
describe emission into a right-going (left-going) mode given by
\begin{equation}\label{SME0}
\mathcal{\tilde{L}}_{\text{R(L)}} \rho_{\text{s}} = \pi g^{2} \big( 2 \sigma_{-} \rho_{\text{s}} \sigma_{+} -  \sigma_{+} \sigma_{-} \rho_{\text{s}} -  \rho_{\text{s}} \sigma_{+} \sigma_{-} \big).
\end{equation}
It follows that the decay rates are $\Gamma_{\text{R}} = \Gamma_{\text{L}} = 2 \pi g^{2}$. 

With the jump operators $J_{\text{R}}^{-}$ and $J_{\text{L}}^{-}$ defined in Eq.\,\eqref{JumpO}, we can rewrite the master equation \eqref{SME} into a completely equivalent Lindblad form 
given in Eqs.\,\eqref{ME1} and \eqref{ME}.
%of $J_{\text{R}}^{-}$ and $J_{\text{L}}^{-}$ as shown in Eq.\,\eqref{ME} 
Comparison between \eqref{SME} and Eq.\,\eqref{ME} shows that $J_{\text{R}}^{-}$ and $J_{\text{L}}^{-}$ become new jump operators and, in addition, the effective driving strength in the unitary evolution changes from $\alpha$ to $\alpha/2$.

%%%%-------Two Qubits-------

For two qubits the input-output relation can be shown to be 
\cite{LalumierePRA2013}
\begin{equation}\label{SIOrelation}
\begin{split}
r_{\text{out}}(t) &= r_{\text{in}}(t) - i \sqrt{2\pi} g_{1} \sigma_{-}^{(1)}(t) -i \sqrt{2\pi} g_{2} \sigma_{-}^{(2)}(t+\Delta t), \\
l_{\text{out}}(t) &= l_{\text{in}}(t) - i \sqrt{2\pi} g_{1} \sigma_{-}^{(1)}(t) -i \sqrt{2\pi} g_{2} \sigma_{-}^{(2)}(t-\Delta t),
\end{split}
\end{equation}
where $\Delta t$ is the time delay between the two qubits and $g_{i}$ the coupling strength between the $i$-th qubit and the waveguide. To deal with this time delay, we apply the Markov approximation \cite{LalumierePRA2013} by setting $\sigma_{-}^{(i)}(t-\Delta t)=\sigma_{-}^{(i)}(t) e^{i\omega_{\text{eg}}^{(i)} \Delta t}$. In view of the input-output relation \eqref{SIOrelation}, we then define the jump operators as $J_{\text{R}}^{-}=i\,r_{\text{out}}(-\Delta t)$ and $J_{\text{L}}^{-}=i\,l_{\text{out}}(0)$.\\

With these jump operators, we can write down a master equation of Lindblad form:
\begin{widetext}
\begin{equation}\label{QJMaster}
\begin{split}
\frac{d}{dt} \rho_{\text{S}} = i \Bigg[ \rho_{\text{S}}&, \bigg(  \frac{\Delta_{1}}{2} \sigma_{z}^{(1)} + \frac{\Delta_{2}}{2} \sigma_{z}^{(2)} + \big(1-\frac{1}{2} e^{i \Delta_{1} \Delta t} \big) g_{1} \alpha_{k}^{\star} \sigma_{-}^{(1)} +  \frac{1}{2} e^{-i k \Delta t} g_{2} \alpha_{k}^{\star} \sigma_{-}^{(2)} + h.c.  \\
&+\pi g_{1} g_{2} \big( i \sigma_{+}^{(2)} \sigma_{-}^{(1)} e^{-i \omega_{\text{eg}}^{(2)} \Delta t} +i \sigma_{+}^{(2)} \sigma_{-}^{(1)} e^{-i \omega_{\text{eg}}^{(1)} \Delta t}  + h.c. \big) \bigg) \Bigg] + \sum_{i=R, L} J_{\text{i}}^{-} \rho_{\text{S}} J_{\text{i}}^{+} - \frac{1}{2} \big\{\rho_{\text{S}}, J_{\text{i}}^{+} J_{\text{i}}^{-} \big\}. 
\end{split}
\end{equation}
\end{widetext}
It is worth noting that the effective coupling/driving strength on qubit $1$ can be tuned by the detuning and separation with the phase factor $\Delta_{1} \Delta t $. When there is no detuning or separation, the driving of the two qubits is the same. Effective interactions between the two qubits mediated by propagating photons appear in the master equation as expected.

%%%%%%%%------------------Quantum Jump Simulation-------------------------------%%%%%%%%%%%%%%%%%%%%%%%%%%
\section{Quantum Jump Simulation}
\label{Appenddix:QJ}

We use the single qubit case as an example. The two-qubit case is a straightforward generalization. From the master equation given in Eq.\,\eqref{ME} of the main text, we define the non-Hermitian effective Hamiltonian $H_{\text{eff}}$ as
\begin{equation}
\begin{split}
H_{\text{eff}} &= \frac{1}{2} \Delta \sigma_{z} + ( \frac{1}{2} g \alpha_{k} \sigma_{+} + \frac{1}{2} g \alpha_{k}^{\star} \sigma_{-} ) - i \frac{1}{2} \sum_{i=R,L} J_{\text{i}}^{+} J_{\text{i}}^{-} .
\end{split}
\end{equation}
Then the quantum jump simulation follows the standard method  \cite{PlenioRMP1998} 
with $J_{\text{R}}^{-}$ and $J_{\text{L}}^{-}$ as jump operators into the right and left-going modes respectively.

{\it1.} Choose the initial state of the wave function $\ket{\psi(0)}$ by
\begin{equation}
\rho_{\text{s}}(0) = \sum_{\psi} P_{\psi} \ket{\psi (0)} \bra{\psi (0)}.
\end{equation}

{\it2.} Choose a suitable time step $dt$. Note that $dt$ must be smaller than both the lifetime $1/\Gamma = 1/ (4 \pi g^2)$ and the average spacing of photons in the input coherent state $2 \pi/\alpha^{2}$.

{\it 3.} In the absence of a quantum jump, the wave function $\ket{\psi(t)}$ evolves into an unnormalized wavefunction $\ket{\tilde{\psi}(t+dt)} $ according to 
\begin{equation}
\ket{\tilde{\psi}(t+dt)} = (1- i H_{\text{eff}} dt) \ket{\psi(t)}.
\end{equation} 
Then
\begin{equation}
\begin{split}
| \ket{ \tilde{\psi}(t+dt) }|^2 &= 1- i dt \bra{\psi(t)} (H_{\text{eff}} - H_{\text{eff}}^{\dagger}) \ket{\psi(t)} + O(dt^2)\\
          									&= 1 - P_{\text{R}} - P_{\text{L}} \equiv 1 - P,
\end{split}
\end{equation}
where
\begin{equation}
P_{\text{R(L)}} = dt  \braket{\psi(t) | J_{\text{R(L)}}^{+} J_{\text{R(L)}}^{-} | \psi(t) }
\end{equation}
is the the probability of decaying into right-going (left-going) mode and 
%\begin{equation}
$P = P_{\text{R}} + P_{\text{L}}$ 
%\end{equation}
is the total decay probability.

{\it 4.} Choose a random number $r$ between $0$ and $1$. If $r < P_{\text{R}}$, there is a decay into right-going mode:
\begin{equation}
\ket{\psi (t+dt)} = \frac{ J_{\text{R}}^{-} \ket{\psi (t)} }{ \sqrt{ \braket{\psi (t) | J_{\text{R}}^{+} J_{\text{R}}^{-} | \psi (t) } } } = \frac{ J_{\text{R}}^{-} \ket{\psi (t)} }{ \sqrt{P_{\text{R}}/dt  } };
\end{equation}

If $P_{\text{R}} < r < (P_{\text{R}}+P_{\text{L}})$, there is a decay into left-going mode:
\begin{equation}
\ket{\psi (t+dt)} = \frac{ J_{\text{L}}^{-} \ket{\psi (t)} }{ \sqrt{ \braket{\psi (t) | J_{\text{L}}^{+} J_{\text{L}}^{-} | \psi (t) } } } = \frac{ J_{\text{L}}^{-} \ket{\psi (t)} }{ \sqrt{P_{\text{L}}/dt   } };
\end{equation}

If $r>(P_{\text{R}}+P_{\text{L}})$, there is no decay:
\begin{equation}
\ket{\psi(t+dt)} =\frac{ \ket{\tilde{\psi}(t+dt)} }{ \sqrt{1 - (P_{\text{R}}+P_{\text{L}})} }.
\end{equation} 

{\it 5.} The density matrix is the average of all the trajectories $\ket{\psi_{m}}$ obtained from each simulation:
\begin{equation}
\rho_{\text{s}}(t) = \sum_{m} \ket{\psi_{m}(t)} \bra{\psi_{m}(t)}.
\end{equation}

%%%%%%%%------------------Calculation of g2-------------------------------%%%%%%%%%%%%%%%%%%%%%%%%%%

\section{Second-Order Correlation Function}
\label{Appenddix:g2}

\begin{figure}[h]
\center
\includegraphics[scale=0.35]{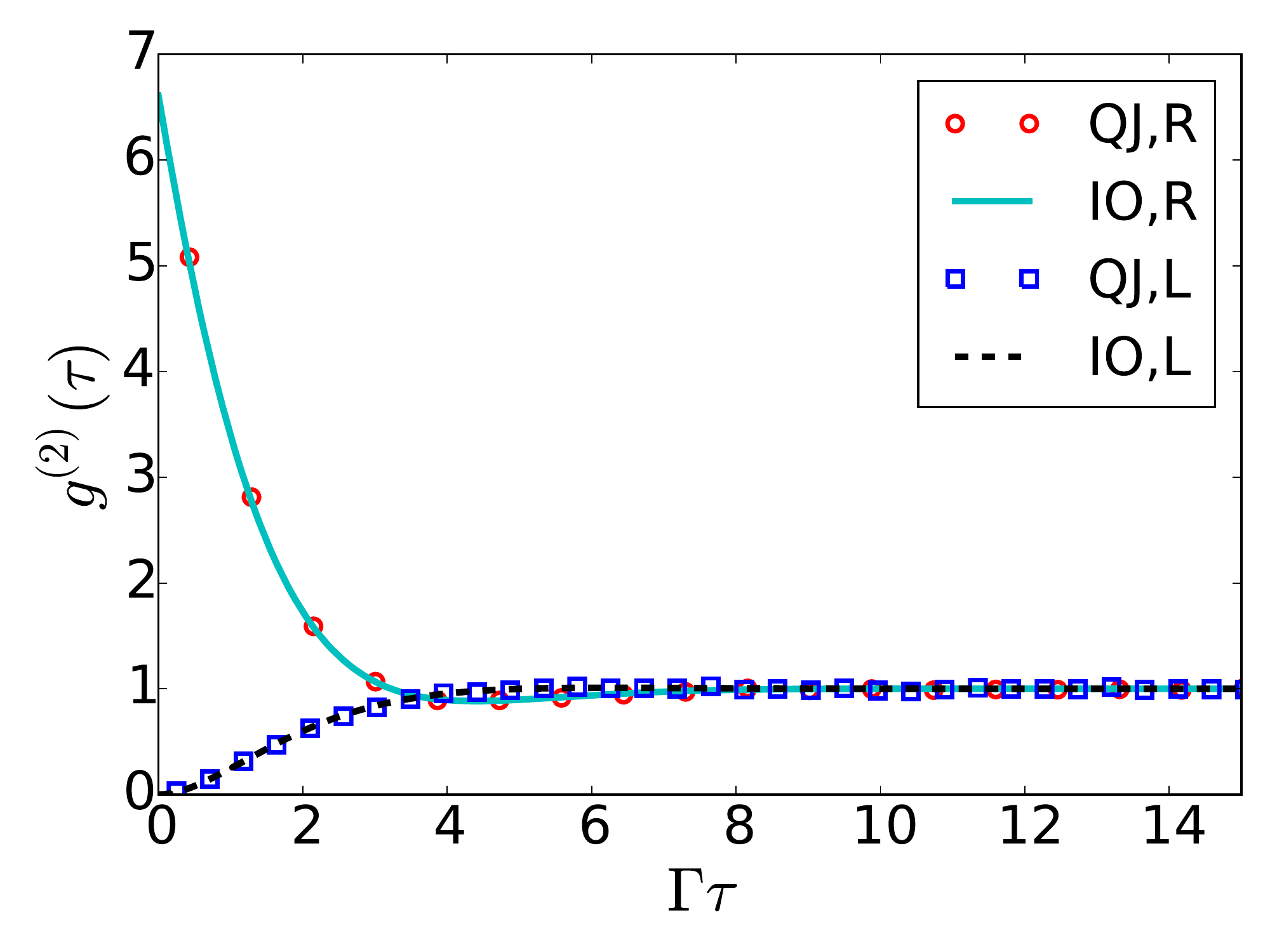}\label{g2}
\caption{Steady state second-order correlation function $g^{(2)}(\tau)$ for transmitted (R) and reflected (L) light for one qubit coupled to a 1D waveguide. Dots are data obtained from the quantum jump simulation (QJ) and lines are results calculated with input-output theory (IO). (Parameters: $\alpha=1$, $\Delta =0$.)}
\end{figure}

As a check of our calculation, the second-order correlation function $g^{(2)}(\tau)$  can be obtained from a quantum jump simulation by plotting a histogram of $\tau$ for many trajectories. 
First choose an initial jump time $t$ in a trajectory. Then record the time intervals $s_{1}, s_{2}, s_{3}, \cdots $ between this jump and all others in this trajectory. Note that there is no `next-photon' requirement for these intervals. Repeat this process for many trajectories, or for the steady state one can use just a single trajectory and vary the starting jump time $t$. The histogram of all these intervals $s_{1}, s_{2}, s_{3}, \cdots $ gives us the $g^{(2)} (\tau)$ as shown in Fig.\,\ref{g2}. It can be seen that the results obtained by this new quantum jump approach agree with the results from input-output theory, within the statistical fluctuations. This verifies that our jump operators correctly describe photon detections at the output ends.

%\begin{figure}[h]
%\center
%\includegraphics[scale=0.36]{Fig6g2.pdf}\label{g2}
%\caption{Steady state second-order correlation function $g^{(2)}(\tau)$ for transmitted and reflected light for one qubit coupled to a 1D waveguide. Dots are data obtained from the quantum jump simulation and solid lines are results calculated with input-output theory. (Parameters: $\alpha=1$, $\Delta =0$.)}
%\end{figure}

\vspace*{-0.45in}
%\newpage
\bibliography{QJ,WQED_2018a.bib}

\end{document}